\documentclass[
 reprint,
 amsmath,amssymb,
 aps,
]{revtex4-1}

\usepackage{graphicx}
\usepackage{dcolumn}
\usepackage{bm}
\usepackage{lipsum}

\newcommand{\mv}[1]{\ensuremath{\mathbf{#1}}} 
\newcommand{\gv}[1]{\ensuremath{\mbox{\boldmath$ #1 $}}} 
\newcommand{\abs}[1]{\left| #1 \right|} 
\newcommand{\avg}[1]{\left \langle #1 \right \rangle} 
\newcommand{\der}[2]{\frac{d #1}{d #2}} 
\newcommand{\pd}[2]{\frac{\partial #1}{\partial #2}} 

\usepackage[colorlinks]{hyperref}  
\usepackage{cleveref}
\hypersetup{linkcolor=blue,citecolor=blue,urlcolor=blue}

\usepackage{titlesec}
\titleformat{\section}
  {\normalfont\fontsize{11}{15}\bfseries}{\thesection}{1em}{}

\usepackage[toc]{appendix}

\crefname{figure}{fig.}{figs.}
\Crefname{figure}{Figures}{Figures}

\begin{document}

\setcounter{page}{1} 

\title{Enhanced persistence and collective migration \\ in cooperatively aligning cell clusters}

\author{Vincent E. Debets$^{1,2}$, Liesbeth M.C. Janssen$^{1,2*}$, Cornelis Storm$^{1,2*}$}

\affiliation{$^{1}$Department of Applied Physics, Eindhoven University of Technology, P.O. Box 513,
5600 MB Eindhoven, The Netherlands\\$^{2}$Institute for Complex
Molecular Systems, Eindhoven University of Technology, P.O. Box 513,
5600 MB Eindhoven, The Netherlands\\}
\email{l.m.c.janssen@tue.nl, c.storm@tue.nl}

\begin{abstract}%
{\noindent Most cells possess the capacity to locomote. Alone or collectively, this allows them to adapt, to rearrange, and to explore their surroundings. The biophysical characterization of such motile processes, in health and disease, has so far focused mostly on two limiting cases: single-cell motility on the one hand, and the dynamics of confluent tissues such as the epithelium on the other. The in-between regime of clusters, composed of relatively few cells, moving as a coherent unit has received less attention. Such small clusters are, however, deeply relevant in development but also in cancer metastasis. In this work, we use cellular Potts models and analytical active matter theory to understand how the motility of small cell clusters changes with $N$, the number of cells in the cluster. Modeling and theory reveal our two main findings: Cluster persistence time increases with $N$ while the intrinsic diffusivity decreases with $N$. We discuss a number of settings in which the motile properties of more complex clusters can be analytically understood, revealing that the focusing effects of small-scale cooperation and cell-cell alignment can overcome the increased bulkiness and internal disorder of multicellular clusters to enhance overall migrational efficacy. We demonstrate this enhancement for small-cluster collective durotaxis, which is shown to proceed more effectively than for single cells. Our results may provide some novel insights into the connection between single-cell and large-scale collective motion and may point the way to the biophysical origins of the enhanced metastatic potential of small tumor cell clusters.}
\end{abstract}

\maketitle 

\section*{Introduction}
\noindent Many cell types---even those that otherwise are largely stationary--- possess an innate capacity to migrate, individually and autonomously, on two-dimensional (2D) substrates or in three-dimensional (3D) matrices. Properly regulated, cell migration contributes crucially to organismal functioning, as it drives vital processes such as morphogenesis, tissue formation, wound healing and the inflammatory response. In pathology, cell migration likewise features prominently, and nowhere more so than in cancer metastasis. Cancer remains one of the leading causes of death in the developed world \cite{Intro} and the vast majority of deaths due to cancer (approximately 90\%) are a consequence of metastasis \cite{Intro1,Intro2,Intro3,Intro4}. In metastasis, cells detach from a primary tumor and invade the surrounding extracellular matrix (ECM)---i.e. the three-dimensional cellularized material that provides structural support to tissue--- migrating towards blood or lymphatic vessels. Once there, migratory cancer cells traverse the vessel wall (intravasation) and pass into the circulation system as circulating tumor cells (CTCs). Eventually, some of these CTCs may once again pass the vessel wall and navigate the local ECM to seed a secondary tumor \cite{Intro2,Intro3,Intro5}. 

The elimination of malignant tumors through early detection and timely resection, possibly combined with chemo-, radiation-, and immune therapy, is the principal directive in treatment. Despite the seemingly straightforward sequence of metastatic events (often referred to as the {\it metastatic cascade}), this process remains poorly understood; effective countermeasures that directly interfere with metastasis itself are scarce \cite{Intro2}. The process continues to hold surprises too: Where it was long generally held that distant metastases were mostly seeded by single tumor cells \cite{Intro6}, recent experimental studies reveal significant contributions to metastasis from so called CTC {\it clusters}: heterogeneous cell clusters consisting of approximately $2$ to $20$ cells that have {\it collectively} detached from a single primary tumor and are {\it collectively} undertaking the entire metastatic cascade; invading, intravasating, and circulating as one conserved unit \cite{Intro1,Intro2,Intro3,Intro5,Intro6,Intro7,Intro8,Intro9}. These clusters are dangerously potent: A study of spontaneous breast cancer in mice revealed that over 97\% of all observed metastases originated from CTC clusters rather than single CTCs. Other work highlights the importance of the cell-cell adhesion mediators such as E-cadherins in metastasis, and likewise suggest that CTC clusters may possess a metastatic potential that is at least $50$ times (and possibly over a $100$ times) greater than for individual CTCs \cite{Intro1,Intro5,Intro7,Intro9,Intro10}. CTC clusters are associated to lowered overall survival and lowered progression-free survival in a range of cancer types \cite{Intro2}. It was convincingly shown that CTC clusters indeed remain a single unit throughout the journey from primary tumor to distant site; the pathway in which polyclonal CTC clusters would assemble from single CTCs at some point during metastasis is highly improbable \cite{Intro5}. Finally, while collective metastasis is our main motivation, we note that collectively moving clusters also play a crucial role in many developmental processes: Refs.\ \cite{cai} and \cite{kolega} emphasize their importance in e.g.\ the neural crest, in mesoblasts in gastrulation, and in the extension of chick somites forming the sclerotome, among a number of other appearances in biology. 

Overall, experimental findings have opened up a completely new field of study focusing on relatively small cell clusters in biology \cite{cai}. In order to explain in particular the enhanced metastatic potential of CTC clusters a number of hypotheses have been brought forward, including the cooperation of heterogeneous cell types within the CTC cluster, shielding from attacks by immune cells, a differential capacity for sensing and responding to chemical gradients \cite{camley, lalli}, and the protection from pressures and shear forces while in the bloodstream \cite{Intro1,Intro6, king}. Yet, much remains unknown about the genesis, transit and the settlement of CTC clusters during metastasis \cite{Intro1}.

The purpose of this article is to examine three physical-mechanical aspects of cluster motility. First: clusters are obviously larger than single cells. How do multiple erratic individual motile tendencies, with varying degrees of coordinated organization, add up to the collective motion of a small cluster of identical cells? Second: How does in-cluster heterogeneity (in intrinsic motility) affect motility at the cluster level? And third: How do these altered properties affect the ability of a cluster to perform durotaxis \cite{Duro1,Duro2,Duro4,Duro5}; that is - to move directedly in the presence of a rigidity gradient? The latter has been shown to improve in large aggregates \cite{Duro1}; how does it play out in smaller clusters?

To address these research questions, we combine coarse-grained simulations with analytical active matter theory. Specifically, we use the cellular Potts model (CPM) to simulate cell (cluster) motion. This model is augmented to capture two important features of collective motility: directional persistence and cell-cell alignment. Directional persistence captures the tendency of individual cells to persist directionally for some amount of time \cite{Duro5,Exp_active_cell1,Exp_active_cell2,ABP3}. It is quantified by a persistence time, which corresponds to the average time it takes a cell to deviate significantly from an initial course. Cell-cell alignment refers to the tendency of densely packed motile cells to mutually inform the direction of their motion \cite{kolega,camleyPRE2014,CPM5}, and has been invoked to explain collective motility in dense systems \cite{Alignment1}. This may happen either by direct physical interactions such as volume exclusion and traction forces where cell-cell adhesions drag neighbors along, or in a more indirect fashion through contact inhibition of locomotion (CIL) \cite{zimmermanPNAS2016}. Although the latter tends to cause cells to move away from each other, in dense systems this effect also suppresses convergent relative motion and is thus generally manifested as a parallel-aligning field. In this work, we model persistence and alignment using a Langevin and Vicsek-type \cite{Vicsek1} approach, respectively. More specifically, we implement persistent migration in two and three dimensions using a Langevin description for the stochastic rotational diffusion of the cells' instantaneous direction of motion. Alignment is implemented in a Vicsek-like feedback mechanism, and quantified by the relative weight assigned to neighbour velocities when updating the velocity of a given cell. To rationalize our CPM simulation results, we also develop an analytical model for finite-sized clusters composed of (aligning) active Brownian particles (ABPs), providing more theoretical insight into the cluster migration efficacy as a function of cluster size and cluster heterogeneity.

The paper is organised as follows. We start with introducing the CPM and demonstrate how a persistent random walk and a Vicsek cell-cell alignment term are implemented in the CPM. We validate the implementation of persistence and alignment by analyzing the trajectories of single cells and cell clusters exploring homogeneous environments. We then discuss the theory of ABPs, and use it to provide an analytical underpinning of the numerically observed behaviors. Finally, we relate the enhanced persistence of clusters to cell transport in a more complex, durotactic environment. We conclude by summarising the main findings and provide some future directions and topics where our results may have an impact.

\section*{Cellular Potts Model}

\begin{figure*}[th!]
\hspace*{-0.5cm}
\includegraphics[scale=0.65]{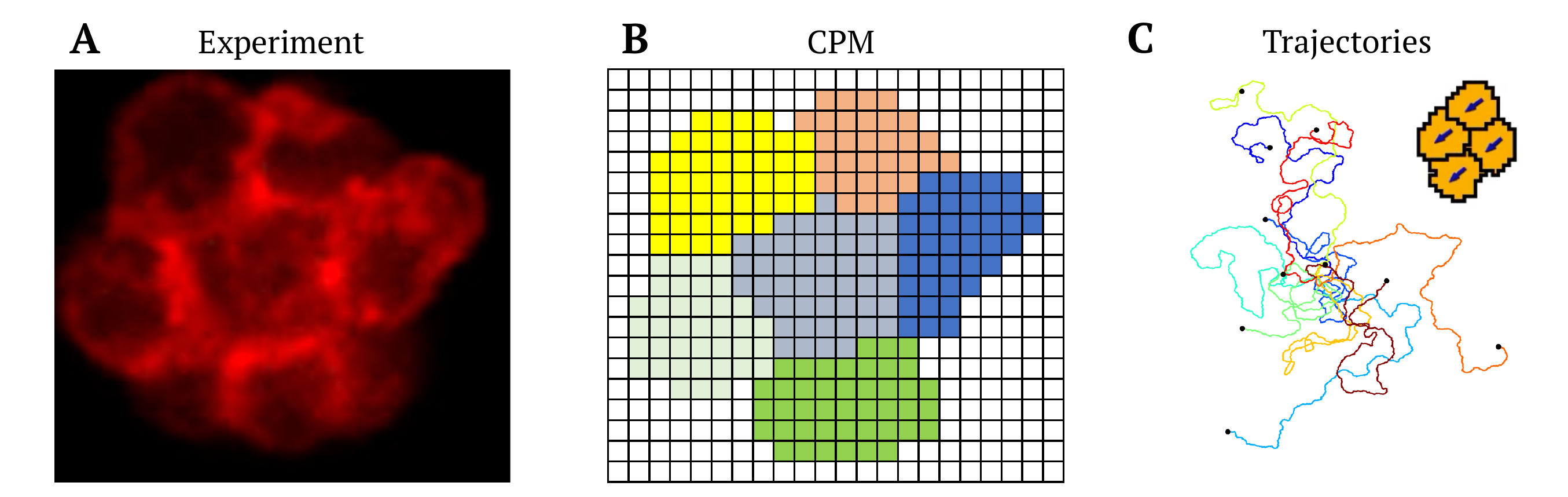}

\caption{(A) Experimental observation of a circulating tumor cell cluster (CTC cluster) taken from \cite{Intro9}. (B) Visualisation in 2D of CPM cells representing the experimentally observed CTC cluster. Different colors represent different spins $\sigma$ with the medium ($\sigma = 0$) shown in white. (C) Example trajectories of an aligned four-cell CPM cluster with the black dots denoting the end point of each trajectory. Inset shows the respective cluster with the arrows denoting the polarity vector $\mv{p}_{\sigma}$ of each cell.}
\label{CPM_schematic}
\end{figure*}

\noindent To simulate the motion of a CTC (cluster) through the ECM we employ the so-called cellular Potts model (CPM) \cite{CPM1,CPM2}. This model, credited for explicitly representing the cell shape, has been successfully applied to a wide variety of biological phenomena involving for instance blood vessel network formation, cancer cell invasion, and collective cell motion \cite{CPM3,CPM4,CPM5,CPM14,CPM15,CPM16}. The CPM is a variation on the classic Potts model \cite{CPM6} and consists of integer spins $\sigma(\mv{x}) \geq 0$ on a discrete square or cubic lattice (with a lattice constant $a$), whose sites are characterised by their position in space $\mv{x}$. Biological cells are represented as (simply connected) domains of equal spin $\sigma(\mv{x}) > 0$, while the medium or ECM is assumed to be homogeneous (to exclusively focus on cell-cell alignment) and depicted by $\sigma=0$ (see \cref{CPM_schematic}). 

Cell movement can then be imposed on the system via a modified Metropolis Monte Carlo algorithm \cite{CPM7,CPM8,CPM9}: a candidate lattice site $i$ is randomly chosen and its spin value $\sigma(\mv{x}_{i})$ is attempted to be changed to a randomly picked adjacent spin value $\sigma(\mv{x}_{j})$. The attempt is (provided all cells remain simply connected) accepted with a Boltzmann probability, i.e.\ $\mathrm{min}(1,e^{-\Delta \mathcal{H}/T})$, where $T$ parameterizes the energy associated with membrane fluctuations \cite{CPM1}. The parameter is suggestively called $T$ to emphasize the temperature-like role it plays in tuning dynamics from quiescent to actively disordered, but we stress that it is not an actual temperature. $\Delta\mathcal{H}$ is the resulting change in a phenomenological Hamiltonian $\mathcal{H}$; the latter accounts for all physically relevant terms, which in the original model are (approximately) constant cell volume and a finite cell-cell interfacial tension yielding a Hamiltonian \cite{CPM1,CPM2}

\begin{equation}\label{Ham}
\begin{split}
    \mathcal{H} & = \mathcal{H}_{\mathrm{volume}} + \mathcal{H}_{\mathrm{adhesion}} \\
    & = \lambda\sum_{\sigma}(V_{\sigma}-V_{\sigma,0})^{2} + \sum_{i,j}\frac{J_{\sigma(\mv{x}_{i}),\sigma(\mv{x}_{j})}(1-\delta_{\sigma(\mv{x}_{i}),\sigma(\mv{x}_{j})})}{\abs{\mv{x}_{i}-\mv{x}_{j}}}.
    \end{split}
\end{equation}
Here $V_{\sigma}$ denotes the volume (in 3D) or area (in 2D) of cell $\sigma$, i.e.\ the number of lattice sites with $\sigma(\mv{x})=\sigma$;  $V_{\sigma,0}$ is the preferred volume or area of the corresponding cell. The parameter $\lambda$ represents the strength of the volume constraint and the first sum is taken over all cell spins $\sigma>0$. For the adhesion term the sum is taken over nearest and next-nearest neighbours $i,j$ with $J_{\sigma,\sigma^{\prime}}\ (= J_{\sigma^{\prime},\sigma})$ denoting the adhesion coefficient between cell $\sigma$ and cell $\sigma^{\prime}$ (or the medium $\sigma^{\prime}=0$), $\abs{\mv{x}_{i}-\mv{x}_{j}}$ the distance between the neighboring sites, and $\delta_{\sigma,\sigma^{\prime}}$ the Kronecker delta which ensures that only lattice site pairs of different cells contribute to the surface energy. Note that generally $J_{\sigma,\sigma^{\prime}}>0$, and that by choosing different values for the coefficient $J$ between two cells and between a cell and the medium we may implement preferential cell-cell adhesion. To quantify the evolution of the system within the CPM we introduce the Monte Carlo step (MCS) as a time measure \cite{CPM1,CPM2,CPM8}. The MCS is defined as $N_{l}$ attempts to change a spin value, with $N_{l}$ the total number of sites in the lattice; it ensures that on average each lattice site is updated once every MCS, thereby decoupling the time step from the actual system size \cite{CPM8}.

\subsection*{Activity \& Persistence} 
\noindent In its original formulation, i.e.\ \cref{Ham}, cell dynamics in the CPM arises solely from fluctuations in the cell volume and interfacial area (or cell area and interfacial length in 2D systems). As a result, the cells do not experience any directional bias. In real life, however, cells migrate actively, and may exhibit biased, directional motion. This may be because of external guiding cues such as the local organization of the extracellular matrix, and more generally  in response to gradients of some kind. In such cases, the motion is called a \textit{taxis}. The most well-known of these tactic motions is chemotaxis, in which cells move upstream in gradients of beneficial compounds such as nutrients or oxygen. To implement such directed motion, which we assume in one form or other to feature in CTCs migrating through the ECM, an additional energy bias $\Delta\mathcal{H}_{a}$ is incorporated in the change of the Hamiltonian $\Delta\mathcal{H}$. This bias promotes attempts that move the cell along a preferred direction which we shall call the \textit{polarization}, and is given by \cite{CPM5,CPM9,CPM10}

\begin{equation}\label{activeH}
    \Delta\mathcal{H}_{a} = -\sum_{\sigma=\sigma(\mv{x}_{i}),\sigma(\mv{x}_{j})} \kappa_{\sigma} \Delta\mv{R}_{\sigma}(\sigma(\mv{x}_{i})\rightarrow \sigma(\mv{x}_{j})) \cdot \mv{p}_{\sigma}.
\end{equation}
Here, $\mv{p}_{\sigma}$ denotes the (unit) polarization vector of cell $\sigma$, i.e.\ the direction in which the cell is currently moving. $\Delta\mv{R}_{\sigma}(\sigma(\mv{x}_{i})\rightarrow \sigma(\mv{x}_{j}))$ is the center-of-mass displacement of cell $\sigma$ that would result if the proposed move were accepted, and $\kappa_{\sigma} > 0$ measures the relative strength of active motion; this parameter controls the speed of cell $\sigma$.

Isolated cells in experiments generally exhibit persistent motion. That is, the direction of motion drifts on some characteristic timescale. Indeed, it has been shown that single cell motility in 2D can be accurately described by a persistent random walk (PRW) \cite{Exp_active_cell1,Exp_active_cell2}, and that consequently its mean square displacement (MSD) is given by \cite{ABP2,ABP3,ABP4,ABP6}
\begin{equation}\label{MSD2D}
    \avg{(\mv{r}(t)-\mv{r}(0))^{2}} = 4D_{a}\tau(e^{-t/\tau} + t/\tau - 1) + 4D_{t}t,
\end{equation}
where $\mv{r}(t)$ is the position of a cell at time $t$. The displacement of the cell is comprised of two parts; a purely diffusive part characterized by a passive diffusion coefficient $D_{t}$, and a persistent contribution quantified by a persistence time $\tau$ and an active diffusion coefficient $D_{a}\equiv v_{0}^{2}\tau/2$ (with $v_{0}$ the active cell speed). At very short times $t \ll \tau$, the resultant motion is diffusive (MSD $\propto t$) with diffusion coefficient $D_{t}$, ballistic (MSD $\propto t^{2}$) at intermediate time scales $t \approx \tau$, and diffusive again with an enhanced diffusion coefficient $D_{t} + D_{a}$ at long times $t\gg \tau$. 

Although the PRW accurately describes cell motility in 2D, the correct description in 3D involves an anisotropic persistent random walk model where two persistent random walks for a primary and nonprimary direction of motion of the environment are combined \cite{Exp_active_cell2}. To make our general point, and to facilitate comparison with an analytical model we will present later on in this paper, we mostly restrict our simulations to 2D cell (cluster) motion. Nonetheless, we note that an extension to 3D leads to similar results (see \cref{AppA}). Interpreting our polarity vector as the instantaneous direction of motion of our active cell, we impose a PRW by letting $\mv{p}_\sigma$ undergo rotational diffusion. This is implemented by expressing it in terms of its polar angle $\phi_{\sigma}$: $\mv{p}_{\sigma} = [\cos(\phi_{\sigma}), \sin(\phi_{\sigma})]$, and having it evolve in time according to a discretised angular Langevin Dynamics process \cite{ABP2,ABP3,ABP4,ABP5,ABP6}:
\begin{equation}\label{Discrete1}
    \phi_{\sigma}(t+\Delta t) = \phi_{\sigma}(t) + \sqrt{\frac{2}{\tau_{\sigma}}} \Gamma(t). 
\end{equation}
Here $\Delta t$ is the time step of the update which we set to 1 MCS, $\tau_{\sigma}$ is the implemented persistence time of cell $\sigma$ (given in units of MCS), and $\Gamma (t)$ is a stochastic white noise term with zero mean, $\langle \Gamma(t)\rangle =0$, and a variance equal to $\Delta t$; $\langle \Gamma(t)\Gamma(t')\rangle =\Delta t\, \delta(t-t')$.

\subsection*{Vicsek Alignment}
\noindent With the update scheme given by \cref{Discrete1}, we have incorporated the persistent random walk into the CPM through reorientations of the polarity vector $\mv{p}_{\sigma}$. This vector represents the currently preferred direction of motion, and may be interpreted as an internal polarization of the motile machinery, i.e.\ the instantaneous polarization direction of cytoskeletal stress fibers or, in a more pragmatic sense, as the orientation of the leading edge of the cell \cite{CPM5} (even though the direction of movement does not always line up perfectly with either of these two directions). For now we treat $\mv{p}_{\sigma}$ as a proxy for some internal or external bias direction that guides the motion.

\begin{figure}[t!]
\hspace*{0.0cm}
\includegraphics[scale=0.38]{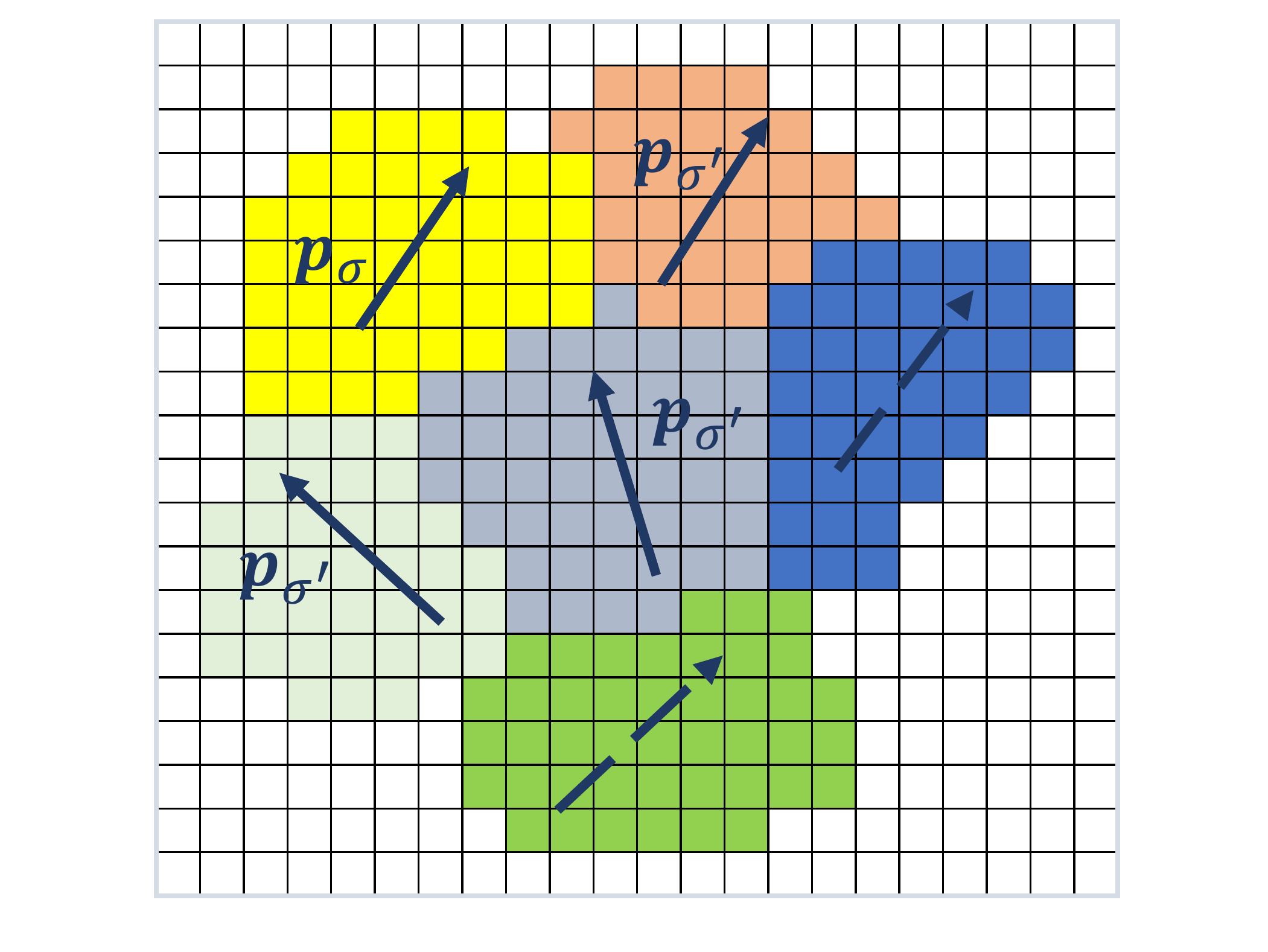}

\caption{Visualisation of the polarity vector $\mv{p}_{\sigma}$ of a CPM cell $\sigma$ (colored in yellow) and its direct neighbor polarities $\mv{p}_{\sigma^{\prime}}$ with which cell $\sigma$ aligns according to a Vicsek-type model. Dashed arrows denote the polarity vectors of the cells in the cluster that are not in direct contact with cell $\sigma$.}
\label{schematic_Vicsek}
\end{figure}

This brings us to a principal feature of this work: the effect of cell-cell  alignment. As detailed in the introduction, it is reasonable to assume that when cells are in contact with each other, they may influence each others's direction of motion and thereby alter the polarity vector $\mv{p}_{\sigma}$ of nearby cells. Inspired by the capacity for neighbor-induced migration alignment in the context of, for instance, wound healing \cite{Alignment1}, we let this interaction between cells manifest itself as a tendency to align their respective polarities in parallel fashion. This is implemented numerically using an adaptation of the well-known Vicsek model \cite{Vicsek1,Vicsek2}, extending the update rule presented in \cref{Discrete1} to
\begin{equation}\label{Discrete2}
    \phi_{\sigma}(t+\Delta t) = \arg \left( \gamma \mv{p}_{\sigma}(t)+\sum_{\sigma^{\prime}} \mv{p}_{\sigma^{\prime}}(t)\right) + \sqrt{\frac{2}{\tau_{\sigma}}} \Gamma(t), 
\end{equation}
where $\gamma$ is a weight factor that controls the degree of alignment, $\arg(\mv{a})$ denotes the angle of a vector $\mv{a}$ in polar coordinates and the sum is taken over all cells $\sigma^{\prime}$ that are in direct contact with cell $\sigma$. We shall call two cells in direct contact when there is at least one site with spin $\sigma$ on the one cell that shares a boundary with the lattice site of a nearby cell $\sigma^{\prime}$ (see \cref{schematic_Vicsek}). Note that this definition of neighbourhood slightly deviates from the original formulation of the Vicsek model, which aligns all particles within an interaction radius. When $\gamma \rightarrow \infty$ the alignment disappears (all direct contacts carry zero relative weight in the update scheme, recovering \cref{Discrete1}), whereas for $\gamma=1$ the polarity vector $\mv{p}_{\sigma}$ instantaneously takes on the average direction of itself and its neighbors after each update and we have perfect local alignment. We will call this the {\it fast-aligning} regime, and it is also how the alignment in the original Vicsek model is implemented \cite{Vicsek1,Vicsek2}.

\subsection*{Simulation Details}
\noindent Each simulation starts with initiating a model CTC cluster by placing $N_{\mathrm{cells}}$ square cells---domains of equal size and each with a unique spin $\sigma>0$---adjacent to each other on a square lattice with grid size $a$. The system is then equilibrated by running the CPM simulation for $500$ MCS including only the original Hamiltonian \cref{Ham}, without the active energy bias \cref{activeH}. This is done to allow the cells to relax to a natural, smoothly convex shape. After this equilibration stage we assign polarity vectors drawn from a uniform distribution to each cell, set the cluster center of mass to $\mv{R}_{c}(t_{0}) \equiv 0$ (which defines the origin), and start the clock at $t_{0}=0$. We then run the the actual simulation using the Hamiltonian \cref{Ham} including the active energy bias \cref{activeH}. We proceed to track the cluster (or single cell) center of mass $\mv{R}_{c}(t_{n})$ at fixed time intervals $\Delta t = t_{n+1}-t_{n}= 1\ \mathrm{MCS}$ to generate the motile trajectory of the cluster (see \cref{CPM_schematic} for example trajectories).

For now we will assume all cell parameters to be spatially independent, and each cell to be identical. That is, we set $\kappa_{\sigma}=\kappa$, $\tau_{\sigma}=\tau$, $V_{\sigma,0}=V_{0}$, $J_{\sigma,\sigma^{\prime}}=J_{\mathrm{cell-cell}}$ for $\sigma,\sigma^{\prime}>0$, and $J_{\sigma,\sigma^{\prime}}=J_{\mathrm{cell-substrate}}$ for $\sigma \vee \sigma^{\prime} = 0$. Additionally, for all 2D simulations in this work we have fixed the simulation pseudo-temperature $T = 1$, the target area of the cells $V_{0} = 64\ a^{2}$, the area constraint strength $\lambda = 1$, the cell-cell line tension $J_{\mathrm{cell-cell}}=0.5$, and the cell-substrate line tension $J_{\mathrm{cell-substrate}}=1$. The positive difference between $J_{\mathrm{cell-substrate}}$ and $J_{\mathrm{cell-cell}}$ implies cells prefer boundaries with other cells over boundaries with the substrate. Furthermore, the fact that both values are individually positive implies that all boundaries experience a positive (contractile) line tension. Thus, effectively, this choice of $J$'s encodes both cell-cell adhesion and cortical tension. Combined with an active energy bias and cell persistence time of typically  $\kappa=5$ and $\tau=500\ \mathrm{MCS}$, respectively, these parameters ensure that cells tend to stick together and, by mapping $a\sim 1\mu$m and $\mathrm{MCS}\sim 0.001$h, single cells have, consistent with experiment, a typical size of $\sim 10\ \mu \mathrm{m}$, a speed of $\sim 50\ \mu \mathrm{m}/\mathrm{h}$, and a persistence time of $\sim 1\ \mathrm{h}$ \cite{Duro1,Duro2,CPM13}.

Finally, to prevent unphysical disintegration of the cell shape as a consequence of strong cell-cell adhesion, we have included an additional shape-regulating contribution to the energy, $\Delta \mathcal{H}_{r}$. This bias term forces the cells to prefer a circular shape, thus penalizing e.g.\ fingering-type structures. In principle, this would be taken care of by the positive cortical tension, but for small systems we find that lattice effects on the shape are non-negligible. The $\Delta \mathcal{H}_{r}$ term may be physically interpreted as a bending rigidity of the cell cortex/perimeter, and we have verified that its precise value does not influence our main findings (see \cref{AppB}) for more details).

\section*{Migration in Uniform Environments: Phenomenology}

\subsection*{Single Cell Motion}

\begin{figure*}[ht!]
\centering
\includegraphics[scale=0.46]{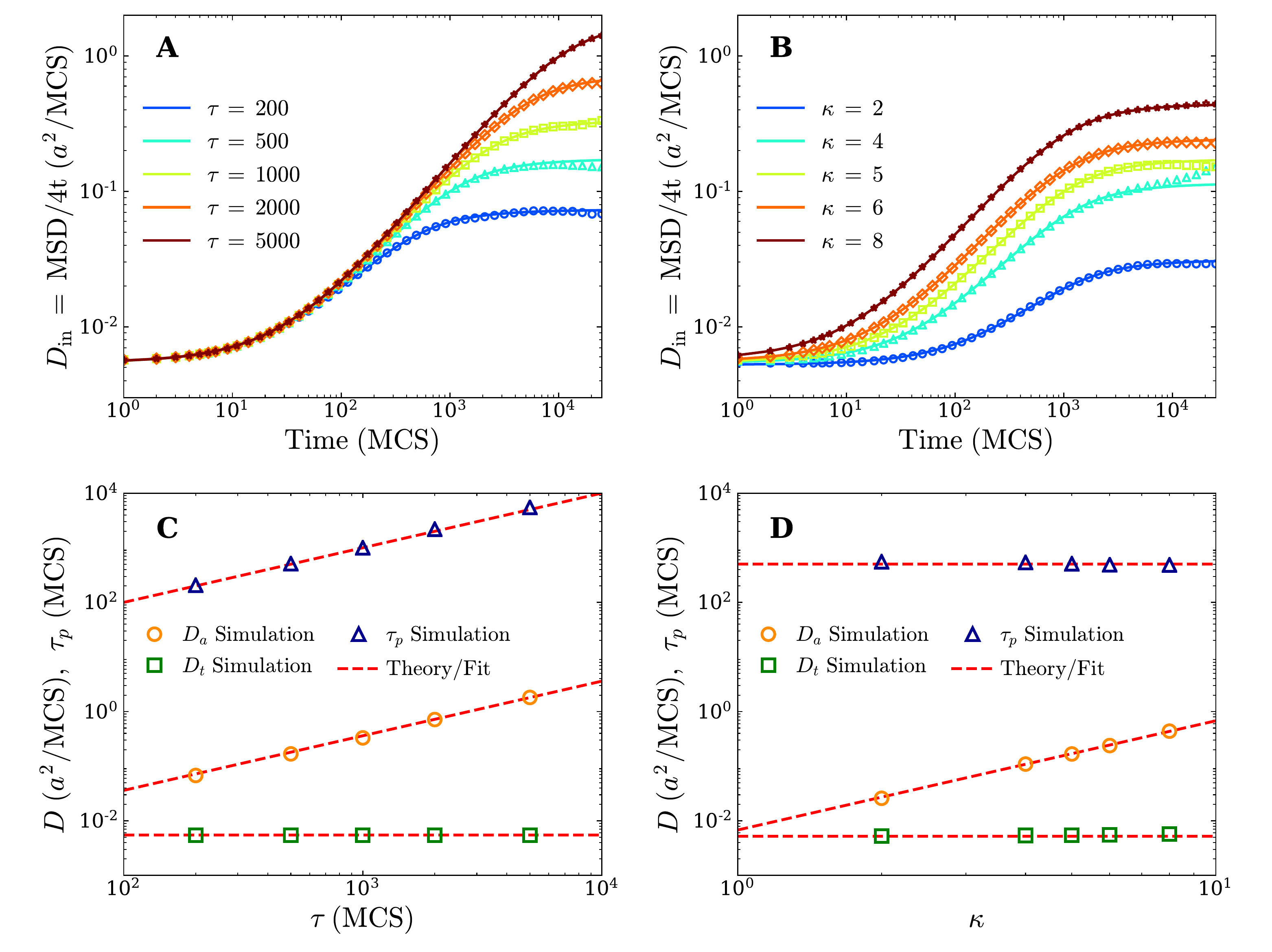}

\caption{(A-B) Plots of the instant diffusion coefficients $D_{\mathrm{in}} = \mathrm{MSD}/4t$ (markers) which have been calculated from 2D single cell CPM simulations and are fitted with a PRW using \cref{MSD2D} (lines). Results are obtained for (A) different implemented persistence times $\tau$ with a fixed value of $\kappa=5$ and (B) different polarity strengths $\kappa$ with a fixed value of $\tau = 500\ \mathrm{MCS}$. (C-D) Persistence time $\tau_{p}$, active diffusion coefficient $D_{a}$, and thermal diffusion coefficient $D_{t}$ obtained from the fits shown in (A-B) respectively. Red dotted lines denote the predicted or fitted values corresponding to $\tau_{p}=\tau$ and $\kappa \propto v_{0}$. Data has been obtained by time-ensemble averaging over $50$ trajectories each consisting of $50000\ \mathrm{MCS}$.}
\label{Single_cell_results}
\end{figure*}

\noindent Before proceeding to the effects of cell-cell alignment in clusters, we first validate our implementation of the PRW into the CPM. For this we study the MSD of isolated cells for different values of $\tau$ and $\kappa$. To analyze the diffusive process we plot the instant diffusion coefficients $D_{\mathrm{in}}=\mathrm{MSD}/4t$ that follow from the calculated MSDs and fit the results with a PRW [\cref{MSD2D}]. This is demonstrated in \cref{Single_cell_results}. The accurate fit confirms that indeed the MSD follows a PRW. By plotting the instant diffusion coefficient we clearly recognize the transition from an initial 'slow' diffusive process (constant $D_{\mathrm{in}}=D_t$) via an intermediate ballistic regime (manifesting as $D_{\mathrm{in}}\sim t$), to again a diffusive process with an increased diffusion coefficient $D_{\mathrm{in}}=D_t+D_a$ in the long time limit.

We further characterise the cell motion from each fitted MSD by extracting the persistence time $\tau_{p}$ (we add the subscript to distinguish the fitted persistence time from the implemented one; we will do this throughout), the active diffusion coefficient $D_{a}$ (or an average active cell speed $v_{0}$), and the 'thermal' diffusion coefficient $D_{t}$. The resulting values of these parameters are shown as a function of both $\tau$ and $\kappa$ in \cref{Single_cell_results}. These results confirm that the persistence time $\tau$ which we implement in the CPM is also the time observed in the simulation, i.e.\ by $\tau_{p}$, and that it is independent of the polarity strength $\kappa$. Also, as anticipated, the active diffusion coefficient $D_{a}$ scales linearly with $\tau$ and quadratically with $\kappa$. Taking into account that for a PRW $D_{a}\propto v_{0}^{2}\tau$, we conclude that $\tau_{p}=\tau$ and $v_{0} \propto \kappa$. This implies that indeed we dial in the persistence time and active speed of individual cells directly with $\tau$ and $\kappa$. The observed values for $D_{t}$ remain constant upon changing both $\tau$ and $\kappa$. This, too, is as expected: The passive motion originates from the pseudo-thermal fluctuations in cell area and shape effected by the parameter $T$. This intrinsic randomness is completely independent of all other parameters.

Thus, consistent with earlier work in e.g. \cite{CPM5} and \cite{CPM9}, we have demonstrated that an active PRW can effectively be mapped onto the CPM. In a broader context than cell motility we note that this implementation also provides a good framework for the study of a larger range of active (soft) materials, and is not necessarily limited to a description of biological cells.

\subsection*{Aligned Cell Cluster Motion}
\noindent We now turn to the collective motion of CTC clusters, focusing specifically on the role of cell-cell alignment and cluster size. Let us first consider the case of fast alignment ($\gamma=1$) for a cluster of $N_{\mathrm{cells}}$ identical cells. The results for $D_{\mathrm{in}}$, extracted from the corresponding MSDs,

\begin{figure*}[ht!]
\includegraphics[scale=0.54]{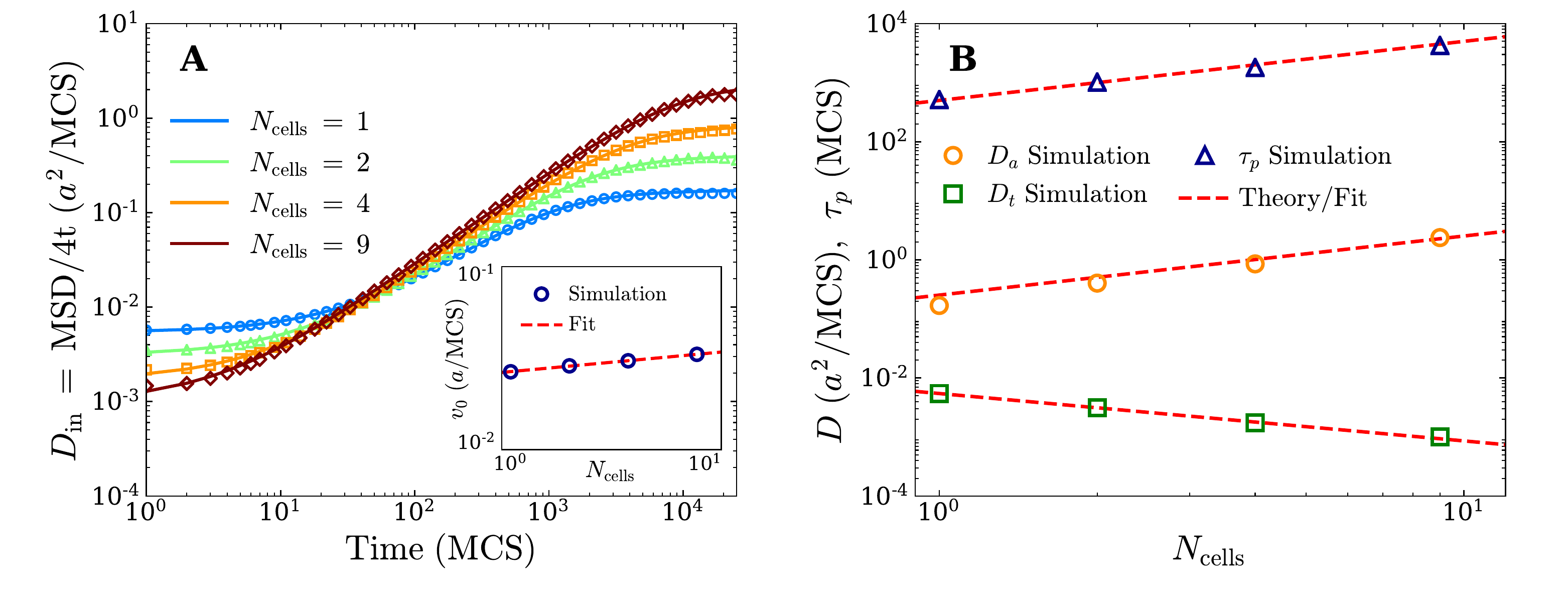}

\caption{(A) Plots of the instant diffusion coefficient $D_{\mathrm{in}} = \mathrm{MSD}/4t$ for fast-aligning ($\gamma=1$) 2D cell clusters consisting of a variable amount of $N_\mathrm{cells}$ identical cells (markers). The results have been fitted with a PRW fitted using \cref{MSD2D} (lines). Inset: average cell (cluster) velocity $v_{0}$ as a function of $N_\mathrm{cells}$ including a power law fit. (B) Cluster persistence time $\tau_{p}$, active diffusion coefficient $D_{a}$, and thermal diffusion coefficient $D_{t}$ obtained from the fits shown in (A). Red dotted lines show a power law fit for $D_{t}$ and a comparison and fit of respectively $\tau_{p}$ and $D_{a}$ to the derived results of the fast-aligning active Brownian particle theory, i.e.\ \cref{Fast_align_res}. Simulation parameters used: $\kappa=5$ and $\tau = 500\ \mathrm{MCS}$. Data has been obtained by time-ensemble averaging over $30$ trajectories each consisting of $50000\ \mathrm{MCS}$.}
\label{Alignment_identical}
\end{figure*}

\noindent are shown in \cref{Alignment_identical}. The MSDs are well-defined, indicating sufficiently large sample sizes, and are still accurately fitted with a PRW; we conclude that an aligning cluster, too, moves according to a PRW. The question, then, is how its parameters depend on cluster size and alignment.

The resulting fit parameters, plotted as a function of $N_{\mathrm{cells}}$ in \cref{Alignment_identical}, allow us to extract these dependencies. We observe that the 'thermal' diffusion coefficient decreases with the number of cells. This can be attributed to the increased size of the cluster that results in weaker relative fluctuations in shape and size. A power law fit yields $D_{t} \sim N_{\mathrm{cells}}^{-0.8}$, showing that this decrease is roughly linear with the number of cells. However, note that these fits (and the subsequent ones) are only made over one decade in $N_{\mathrm{cells}}$ and so the significance of the fitted powers is limited. 

The persistence time of the cluster, on the other hand, is seen to increase linearly with the number of cells: $\tau_{p} \sim N_{\mathrm{cells}}\tau$. An intuitive explanation for this may be found in the fact that when the cells are strongly aligning (sufficiently small $\gamma$), (almost) all cells must simultaneously reorient towards the same direction in order to permit the entire cluster to change its course. This suggests that larger clusters of cells prone to alignment generally continue to move along the same direction for longer times, which corresponds to an increasing persistence time.

Similar to the persistence time, the active diffusion coefficient $D_{a}$ also increases linearly with $N_{\mathrm{cells}}$. We may understand this scaling by noting that for a PRW, the active diffusion coefficient is expected to scale linearly with the cluster persistence time $D_{a}\propto \tau_{p}$ and thus, by extension, also with $N_{\mathrm{cells}}$. This holds exactly when the cluster speed is independent of the cluster size, and remains the same as that of a single isolated cell. We do see, however, that cell clusters actually have a slightly larger active speed than single cells. This is demonstrated in the inset of \cref{Alignment_identical}, which shows the average absolute velocity $v_{0}$ as function of $N_{\mathrm{cells}}$. However, a power law fit yields $v_{0}\propto N_{\mathrm{cells}}^{0.1}$; hence there is only a weak dependence of the cell cluster speed on the number of cells that does not strongly influence the persistent motion of the cluster.

Interestingly, comparable results are obtained for 3D simulations (see \cref{AppA}). This suggests that the effect of the Vicsek alignment of cell polarities on the cell cluster motion is the same in 3D as it is in 2D. In particular, it shows that within a more extended 3D model setup, the effect of alignment still results in a linear dependence of the persistence time on the number of cells in a cluster.

Up until this point, we have imposed fast neighbor alignment within the cluster by setting $\gamma=1$. In order to assess the influence of the relative weight of the neighboring polarizations, we have calculated MSDs for cell clusters experiencing weaker alignment (by setting $\gamma=50$). For this $\gamma$, the MSDs and resulting fit parameters do not change noticeably; we find roughly the same results as for $\gamma=1$ \cite{thesisDebets}. Thus, even moderate degrees of alignment still produce highly cooperative migration in the cluster. There is a finite bound on this effect, however---by increasing $\gamma$ further, the system enters a regime where the alignment is not sufficiently strong anymore and the cluster may even disintegrate. Indeed, our simulations show that when $\gamma$ passes a critical value (roughly, of the numerical order of the persistence time $\tau$), the cluster of cells quickly falls apart into single cells and an analysis of the center-of-mass trajectory becomes meaningless. We can understand this by realising that in the case of no or weak alignment cells often want to travel in different directions (opposite polarity vectors) for long times and can then actively pull themselves loose from the other adjacent cells. Furthermore, the disintegration of the cluster suggests that there exists a critical degree of alignment $\gamma_c$ beyond which alignment is not sufficiently strong to keep the cluster together.

Thus, we have demonstrated that fast alignment of the cells in the CPM will increase the persistence of the cluster, allowing it to move more directionally. This happens at the cost of a decrease in the translational diffusion coefficient. In the case we consider, and which we assume to most closely represent actual cellular behavior, the overall motion is dominated by its active contribution; ($D_{a} \gg D_{t}$). As a result, the decrease in the 'thermal' diffusion coefficient will hardly influence the overall motion. This implies that aligned clusters can, on average, cover more distance than single cells within a given timeframe, provided that \ $v_{0}$ is sufficiently large. In the context of CTC clusters, this might allow them to reach targets such as blood vessels more easily. It also suggests that when clusters experience an externally imposed polarity (through e.g.\ tracks or anisotropy in the ECM), they are generally better able to follow such tracks collectively compared to single cells, enhancing the effects of contact guidance.  

\begin{figure}[ht]
\hspace*{-0.8cm}
\includegraphics[scale=0.5]{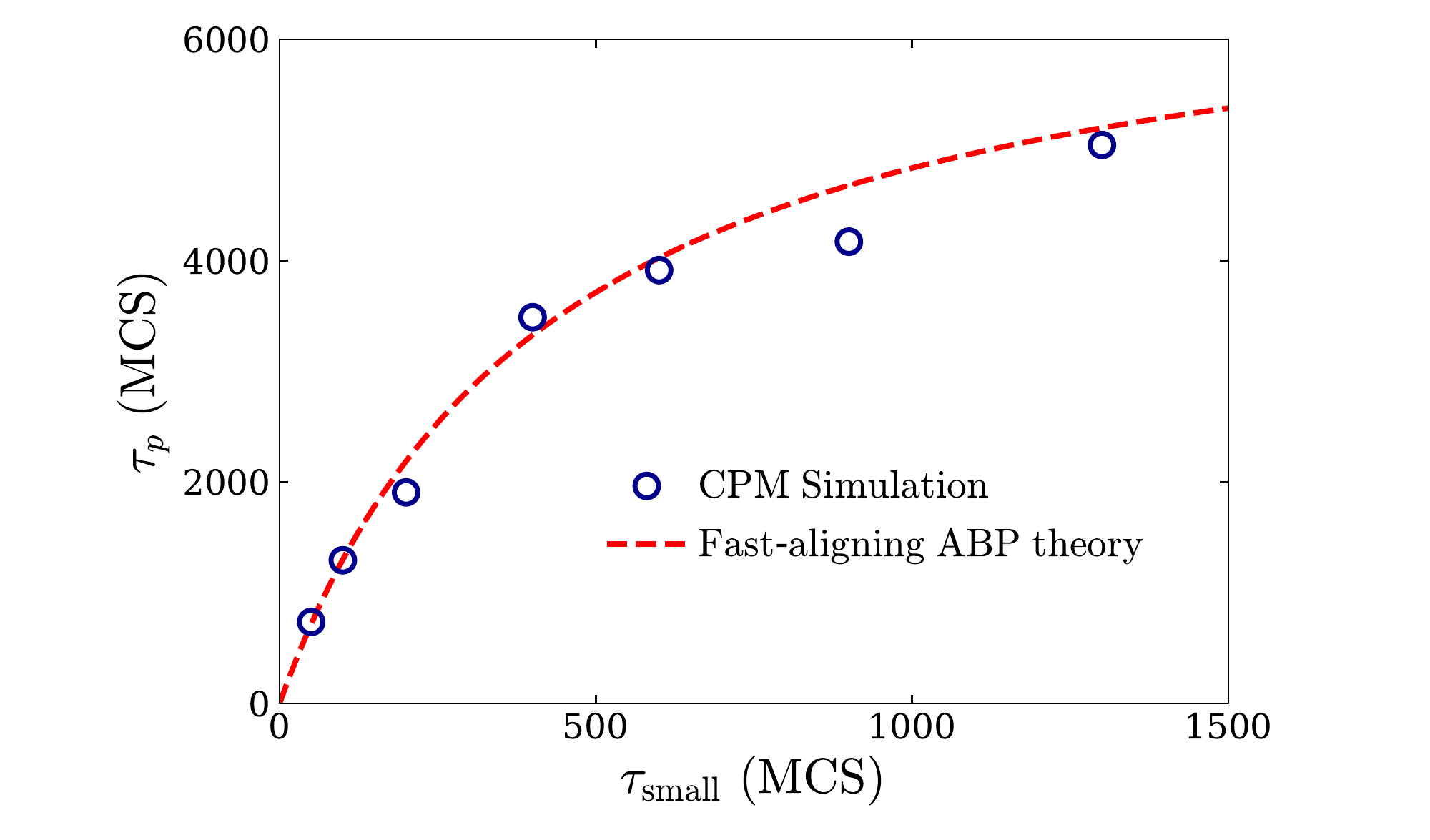}
\caption{Plot of the cluster persistence time $\tau_{p}$ as a function of the implemented small persistence time $\tau_{\mathrm{small}}$. Circles denote the results obtained from PRW fits of the MSD which in turn has been calculated from fast-aligning CPM cell cluster trajectories. Red-dotted line represents the theoretical prediction of our fast-aligning active Brownian particle theory, i.e.\ \cref{Non_iden_taucm}. Simulation parameters used: $N_{\mathrm{cells}}=4$, $\gamma=1$, $\kappa=5$, and $\tau_{\mathrm{large}} = 1300\ \mathrm{MCS}$. Data for the MSDs has been obtained by time-ensemble averaging over $30$ trajectories each consisting of $50000\ \mathrm{MCS}$.}
\label{Non_identical_taup}
\end{figure}

Finally, we consider the effect of fast alignment for a heterogeneous cluster. In particular, we investigate how one less persistent cell influences the motion of an otherwise more persistent cluster. Indeed, individual cells typically show a variety of persistence times in experiments \cite{Exp_active_cell2}, and thus CTC clusters will consist of a heterogeneous mixture of cells. To study this effect, we simulate the motion of a fast-aligning cluster consisting of $N_{\mathrm{cells}}=4$ cells, $3$ of which have a 'large' persistence time (denoted $\tau_{\mathrm{large}}$) of $1300$ MCS, and one has a variable 'small' persistence time (denoted $\tau_{\mathrm{small}}$). As before, we retrieve the cluster persistence time from fitting the collective MSD. The resulting values are plotted as a function of $\tau_{\mathrm{small}}$ in \cref{Non_identical_taup}. It shows that the collective benefit of alignment can become much smaller or even non-existent ($\tau_{p} < \tau_{\mathrm{large}}$) by adding a single cell with a small persistence to an existing cell cluster. We can understand this by realising that a small persistence time corresponds to a rapid reorientation of the cell's polarity. If the reorientation becomes too fast, the cell will drag along other cells towards this polarity as well, which results in a decrease of the cluster persistence. This demonstrates that to exhibit the additional directionality of aligning cluster motion, the spread in individual persistence times should not be too large.

\section*{Active Brownian Motion}
\subsection*{Identical Particles}
\noindent To provide a more general framework for our results, we now seek to rationalize the observed benefits of cell-cell alignment in our CPM simulations using so-called active matter theory. One of the most widely used models in this field \cite{ABP1,ABP2,ABP3,ABP4} is the active Brownian particle (ABP) model. Such particles undergo Brownian motion with a 'thermal' diffusion coefficient $D_{t}$, while they simultaneously self-propel with an absolute speed $v_{0}$ along their orientation axis, called the director $\mv{e}_{i}(t)$. The index $i$ here labels each of the $N$ individual ABPs that, together, form a cluster in our theory. The evolution in time $t$ of the position
$\mv{r}_{i}(t)=[x_{i}(t),y_{i}(t)]$ of each particle $i$ is captured by the stochastic differential Langevin equation \cite{ABP1,ABP2,ABP3,ABP4,ABP5,ABP6}
\begin{equation}\label{Langevin1}
    \der{ \mv{r}_{i}(t) }{t} = v_{0}\mv{e}_{i}(t) + \sqrt{2D_{t}}\gv{\xi}_{i}(t).
\end{equation}
Here $\gv{\xi}_{i}= [\xi_{x_{i}},\xi_{y_{i}}]$ where $\xi_{\alpha}\ (\alpha = x_{i},y_{i})$ represents an independent white noise stochastic process with zero mean, $\avg{\xi_{\alpha}(t)} = 0$, and delta-correlations $\avg{\xi_{\alpha}(t^{\prime})\xi_{\beta}(t)} = \delta (t^{\prime}-t)\delta_{\alpha,\beta}$.

Similar to the polarity vector of each CPM cell $\mv{p}_{\sigma}$, the director of each ABP is parametrised by the polar angle $\phi_{i} (t)\ \in \ [0,2\pi)$, i.e.\  $\mv{e}_{i}(t)=[\cos\phi_{i}(t),\sin\phi_{i}(t)]$. In our aligning ABP model, we assume that $\phi_{i}$ evolves in time not only according to a stochastic rotational diffusion process, but also due to a potential $U$ that encodes velocity alignment,
\begin{equation}\label{rot_diff2}
    \der{ \phi_{i}(t) }{t} = -\eta \pd{U}{\phi_{i}} + \sqrt{\frac{2}{\tau}}\xi_{\phi_{i}}(t),
\end{equation}
with $\eta>0$ denoting a relaxation rate that controls how fast the alignment takes place, $\tau$ the single particle persistence time, and $\xi_{\phi_{i}}$ another independent white noise stochastic process. For the aligning potential $U$ we write
\begin{equation}\label{align_pot}
    U(\{\mv{r}_{i}\},\{\phi_{i}\})= -\sum_{\abs{\mv{r}_{i} - \mv{r}_{j}} < r_{c}} \mu_{ij}\cos(\phi_{i} - \phi_{j}),
\end{equation}
where $\{\mv{r}_{i}\}$, $\{\phi_{i}\}$ denote the set of all $N$ positions and angles respectively, $\mu_{ij}>0$ is a coupling constant which for identical particles simplifies to $\mu_{ij}=\mu$, and the sum is taken over all particle combinations $i,j$ that are within one interaction distance $r_{c}$ from each other. Note that this potential has a minimum when both particles point in the same direction ($\phi_{i}=\phi_{j}$), while it exhibits a maximum when particles are pointing in the opposite direction ($\phi_{i}=\phi_{j}+\pi$). Additionally, it has been demonstrated that in the limit of fast angular relaxation, the introduced continuum description of angular alignment given by \cref{rot_diff2,align_pot} is equivalent to the 2D Vicsek model, i.e.\ \cref{Discrete2}  \cite{vel_align1,vel_align2}. This allows us to draw a direct comparison between the CPM simulations and the obtained theoretical results for aligning ABPs. 

We note that when the alignment between ABPs disappears, i.e.\ $\eta=0$, \cref{Langevin1,rot_diff2} represent a system of non-interacting ABPs. In this case the particles will simply follow a PRW and their individual MSDs are given by \cref{MSD2D} \cite{ABP2,ABP3,ABP4,ABP6}.   

\begin{figure}[t!]
\hspace*{-1.5cm}
\includegraphics[scale=0.33]{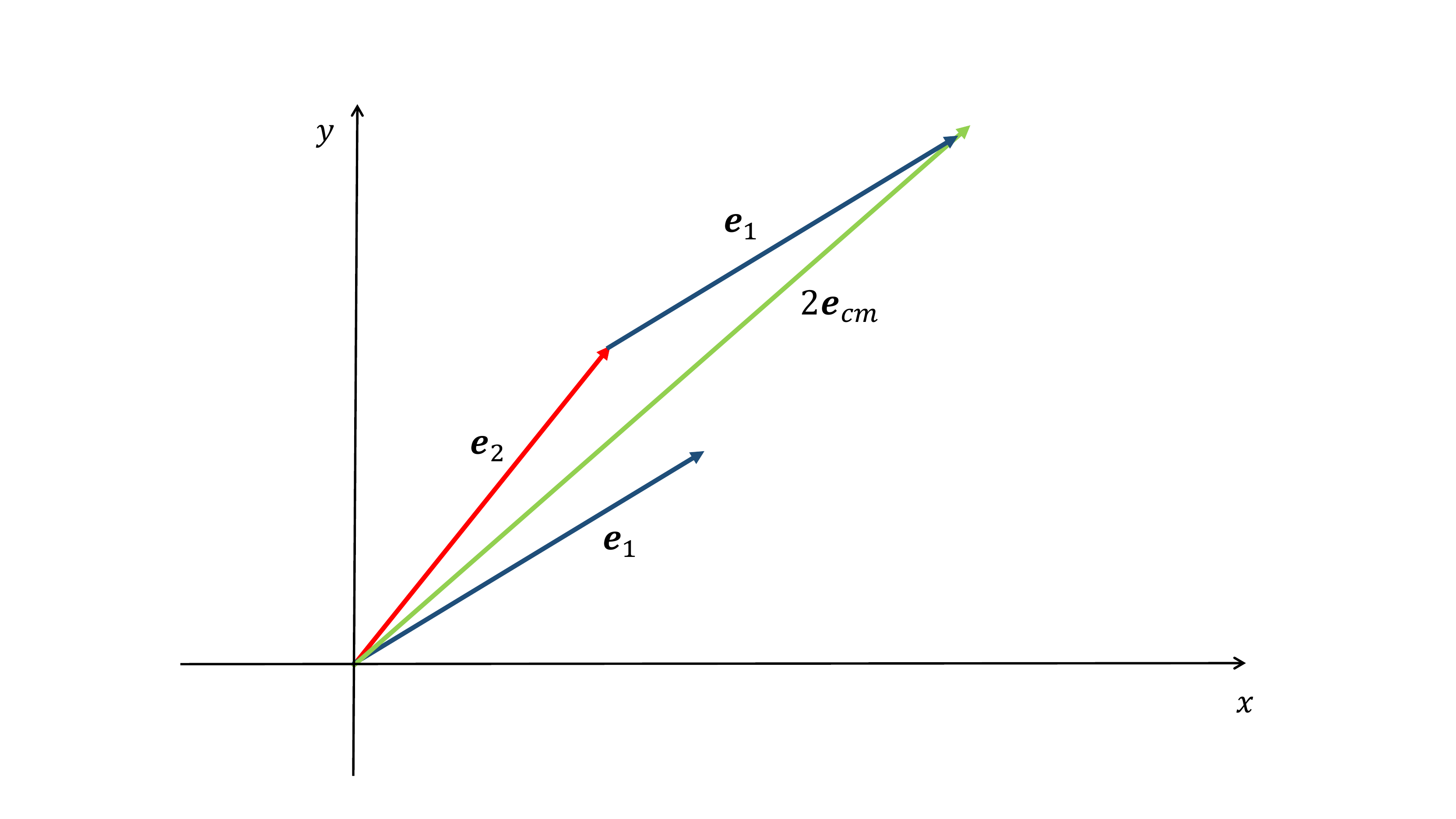}

\caption{Visualisation of the approximation for $N = 2$ particles in which we replace the sum of both particle’s directors $\mv{e}_{1,2} = [\cos(\phi_{1,2}), \sin(\phi_{1,2})]$ by two times the director with the average angle of both. This director is denoted $\mv{e}_{cm} = [\cos(\phi_{cm}), \sin(\phi_{cm})]$ with $\phi_{cm}=(\phi_{1}+\phi_{2})/2$. We have used $(\phi_{2} - \phi_{1}) = \pi/9$; it can be seen that the approximation is still reasonably accurate.}
\label{fast-aligning_approx}
\end{figure}

To describe the motion of the cluster of particles as a whole, we may focus on the center of mass, i.e.\ $\mv{R}=\frac{1}{N}\sum_{i=1}^{N}\mv{r}_{i}$, which obeys the following stochastic differential equation
\begin{equation}\label{Langevin2}
    \der{\mv{R}(t)}{t}=\frac{1}{N}\sum_{i=1}^{N}\left( v_{0}\mv{e}_{i}(t) + \sqrt{2D_{t}} \gv{\xi}_{i}(t) \right).
\end{equation}
To impose fast angular relaxation, we assume that all the particles within the cluster align rapidly with each other ($\mu\eta \gg \tau^{-1}$); hence, the difference between each pair of angles will remain small, i.e. $\abs{\phi_{i}-\phi_{j}}\ll 1$ for all $i,j$. Since the particles travel with equal speeds $v_{0}$, this implies that they remain close together and we expect the cluster of particles (like a CTC cluster) to travel as a whole. Moreover, it allows us to simplify the sums in \cref{Langevin2}. As all involved angles are almost equal, the directors $\mv{e}_{i}$ will point in roughly the same direction. We can therefore approximate the sum of the $N$ directors by $N$ vectors which all point in the average direction of the particles. In other words, we can replace $\sum_{i=1}^{N} \mv{e}_{i} \rightarrow N\mv{e}_{cm}$ in \cref{Langevin2} where $\mv{e}_{cm} = [\cos(\phi_{cm}), \sin(\phi_{cm})]$ and $\phi_{cm} \equiv \frac{1}{N}\sum_{i=1}^{N}\phi_{i}$. A visualisation of this approximation for $N = 2$ is shown in \cref{fast-aligning_approx}. 

Additionally, since the zero mean stochastic processes $\xi_{\alpha}$ are independent and delta-correlated, we can effectively replace a sum of these variables by a single one via $\sum_{i=1}^{N}\xi_{\alpha_{i}} \rightarrow \sqrt{N}\xi_{\alpha_{cm}}$. The factor $\sqrt{N}$ is added to ensure that the correlation remains consistent, i.e.\ $\avg{\sum_{i=1}^{N}\xi_{\alpha_{i}}(t^{\prime})\sum_{i=1}^{N}\xi_{\alpha_{i}}(t)}=N\avg{\xi_{\alpha_{cm}}(t^{\prime})\xi_{\alpha_{cm}}(t)}$. Overall this allows us to simplify \cref{Langevin2} as
\begin{equation}\label{Langevin3}
    \der{\mv{R}(t)}{t}= v_{0}\mv{e}_{cm}(t) + \sqrt{\frac{2D_{t}}{N}} \gv{\xi}_{cm}(t),
\end{equation}
where $\gv{\xi}_{cm} = [\xi_{x_{cm}},\xi_{y_{cm}}]$ represents a new vector of independent stochastic processes with zero mean and delta-correlations. 

The time evolution of the average direction of the particles (and thus of the cluster) can be formulated using \cref{rot_diff2}, i.e.\ 
\begin{equation}
\der{\phi_{cm}(t)}{t} = \frac{1}{N}\sqrt{\frac{2}{\tau}}\sum_{i=1}^{N}\xi_{\phi_{i}}(t),
\end{equation}
where, due to symmetry, all alignment terms cancel against each other. As already mentioned, we can replace a sum of the stochastic noise terms by a single one. Introducing a new stochastic process $\xi_{\phi_{cm}}$ with zero mean and delta-correlations we arrive at
\begin{equation}\label{rot_diff3}
\der{\phi_{cm}(t)}{t} = \sqrt{\frac{2}{N\tau}}\xi_{\phi_{cm}}(t).
\end{equation}

Interestingly, these resulting equations that govern the motion of the center of mass (and thus of the entire cluster of $N$ particles) [\cref{Langevin3,rot_diff3}], are identical in form to the equations that describe a single non-interacting ABP, i.e.\ \cref{Langevin1,rot_diff2} with $\eta=0$. The only difference lies in the fact that the center-of-mass persistence time has increased in proportion to the number of particles $\tau \rightarrow N\tau$, while the 'thermal' diffusion coefficient has decreased as $D_{t} \rightarrow D_{t}/N$. Summarising, we conclude that the center-of-mass motion of a cluster of $N$ fast-aligning ABPs follows a PRW [\cref{MSD2D}] that is characterised by a cluster persistence time $\tau^{cm}$, thermal diffusion coefficient $D_{t}^{cm}$, and active diffusion coefficient $D_{a}^{cm}$ given by 
\begin{equation}\label{Fast_align_res}
\tau^{cm}=N\tau, \quad D_{t}^{cm} =D_{t}/N, \quad D_{a}^{cm} = v_{0}^{2}N\tau/2.    
\end{equation}

Relating our ABPs to the CPM cells by interpreting $N$ as $N_{\mathrm{cells}}$, we see that these theoretical results are in good agreement with the ones from our CPM simulations (see \cref{Alignment_identical}) and thus provide a theoretical underpinning for the increased cluster persistence due to cell (particle) alignment observed earlier. The prediction of the 'thermal' diffusion coefficient scaling as $N^{-1}$ deviates slightly from the observed power of $-0.8$ seen in the CPM simulation. This was to be expected; the passive motion exhibited by a CPM cell is considerably more complex than that of a point particle and in that light it is rather striking that the single, passive diffusive process in the theory so closely resembles the CPM results.

Our analytical argument is not limited to the specific potential we have chosen in  \cref{align_pot}. In addition to the requirement that the particles must align sufficiently quickly (which most likely becomes more difficult for larger $N$), we only require that all alignment contributions to the time evolution of $\phi_{cm}$ will cancel out. In other words, for all potentials that satisfy 
\begin{equation}\label{sum_align}
    \sum_{i=1}^{N}\pd{U}{\phi_{i}} = 0\, ,
\end{equation}
our argument and the results should be valid. 

\subsection*{Non-Identical Particles} 
\noindent To incorporate cluster heterogeneity, i.e.\ to account for the fact that single-cell properties within a CTC cluster are generally not the same \cite{Exp_active_cell2}, we can extend our theory analysis to a set of $N$ quickly aligning non-identical ABPs in several ways. One way would be to let all particles travel at different speeds $v_{0} \rightarrow v_{0,i}$; one could also make the degree of alignment explicitly particle-dependent by substituting $\eta \rightarrow \eta_{i}$ and $\mu \rightarrow \mu_{ij}$ (with $\mu_{ij}=\mu_{ji}$) in \cref{align_pot,rot_diff2} respectively. However, we refrain from exploring these options in too much detail for several reasons. Firstly, letting the particles travel at different speeds will eventually lead to particle separations that are larger than the interaction range of the particles. This scenario is obviously incorrect for CTC clusters in which the cells stick together and travel at approximately equal velocities, and furthermore would invalidate the assumption of fast alignment between particles. Secondly, introducing $\mu_{ij}$ will not change the equations governing the center-of-mass motion: all alignment terms are still canceling out, i.e.\ \cref{sum_align} still applies. Conversely, by introducing a particle-dependent relaxation constant $\eta_{i}$ (e.g.\ to account for different cell sizes with different friction constants), \cref{sum_align} will not be valid anymore. In particular, the time evolution of $\phi_{cm}$ will then also contain terms that are proportional to $(\eta_{i}-\eta_{j})\sin(\phi_{i}-\phi_{j})$. Nonetheless, for strong enough alignment $\abs{\phi_{i}-\phi_{j}}$ will remain sufficiently small, allowing us to neglect these terms and recover the same results as for identical particles. In other words, a larger variety in relaxation constants $\eta_{i}$ will result in a more narrow fast-alignment regime for the cluster, but within this regime it does not qualitatively change its motion.


The most relevant unexplored option for introducing heterogeneity in our aligning ABP model is therefore the scenario we have also studied numerically: to have each particle move with a different persistence time, i.e.\ to replace $\tau \rightarrow \tau_{i}$ in \cref{rot_diff2}. This implies that instead of \cref{rot_diff3} we have
\begin{equation}
    \der{\phi_{cm}(t)}{t} = \left(  \frac{2 \left( \sum_{i=1}^{N}\frac{1}{\tau_{i}} \right)}{N^{2}}\right)^{1/2}\xi_{\phi_{cm}}(t) \equiv \sqrt{\frac{2}{\tau^{cm}}}\xi_{\phi_{cm}}(t),
\end{equation}
where the cluster persistence time is now given by
\begin{equation}\label{taucm}
    \tau^{cm}=N^{2} \left( \sum_{i=1}^{N}\frac{1}{\tau_{i}} \right)^{-1}.
\end{equation}
Note that again we have replaced the sum of stochastic terms by a single one, but due to the particle-dependent $\tau_{i}$ this is less straightforward, i.e.\ $\sum_{i=1}^{N}\sqrt{\frac{1}{\tau_{i}}}\xi_{\alpha_{i}} \rightarrow \left( \sum_{i=1}^{N}\frac{1}{\tau_{i}} \right)^{1/2} \xi_{\alpha_{cm}}$. When all persistence times are equal ($\tau_{i}=\tau$), we recover the linear increase of the persistence time with the number of particles ($\tau^{cm}=N\tau$).

It is now interesting to see how the behavior of a cluster with a distribution of persistence times compares to the case in which all particles are identical. In fact, from \cref{taucm} it can be (straightforwardly) derived (see \cite{thesisDebets} for details) that for each set of $N$ persistence times $\{\tau_{i}\}$ we have 
\begin{equation}
\tau^{cm} \leq N\avg{\tau} 
\end{equation}
with $\avg{\tau}=\frac{1}{N}\sum_{i=1}^{N}\tau_{i}$ the average persistence time of the set and the equal sign corresponding to a constant $\tau_{i}=\tau$. This shows that broadening the distribution of individual persistence times of particles (keeping a constant average) will always decrease the center-of-mass persistence time $\tau^{cm}$. In other words, alignment is always less effective in terms of increasing $\tau^{cm}$ when individual particles travel with a different persistence.

Finally, let us verify one last numerical observation: that even a single less persistent particle among the aligning particles can cancel out the benefits of alignment of the total cluster. Suppose we have $N$ fast aligning particles, $N-1$ of which have an individual persistence time $\tau_{\mathrm{large}}$, and one particle has a smaller persistence time $\tau_{\mathrm{small}} < \tau_{\mathrm{large}}$. The persistence time of the center-of-mass motion is then given by
\begin{equation}\label{Non_iden_taucm}
    \tau^{cm} = \frac{N^{2}\tau_{\mathrm{small}} \tau_{\mathrm{large}}}{(N-1)\tau_{\mathrm{small}} + \tau_{\mathrm{large}}},
\end{equation}
which agrees very well with our CPM simulation results (see \cref{Non_identical_taup}) and confirms that one particle can substantially decrease the mobility benefits of alignment. Particularly, the effect of collective alignment will be canceled when $\tau^{cm}=\tau_{\mathrm{large}}$. In that case, we find
\begin{equation}
    \tau_{\mathrm{small}} = \frac{\tau_{\mathrm{large}}}{N^{2}-N+1}.
\end{equation}
This value for $\tau_{\mathrm{small}}$ thus presents a critical value below which the cluster moves with less persistence than the individual particles. Note that for large $N$ we have $\tau_{\mathrm{small}} \sim \frac{\tau_{\mathrm{large}}}{N^{2}} \rightarrow 0$ and a single particle is not able to disturb the collective motion of a large cluster. However, since $N$ is typically not large for CTC clusters, this effect is not negligible and the inclusion of a rapidly reorienting cell to a CTC cluster can, at least in principle, strongly suppress its directional movement. 

\section*{Durotaxis}

\begin{figure*}[ht]
\hspace*{-0.2cm}
\includegraphics[scale=0.44]{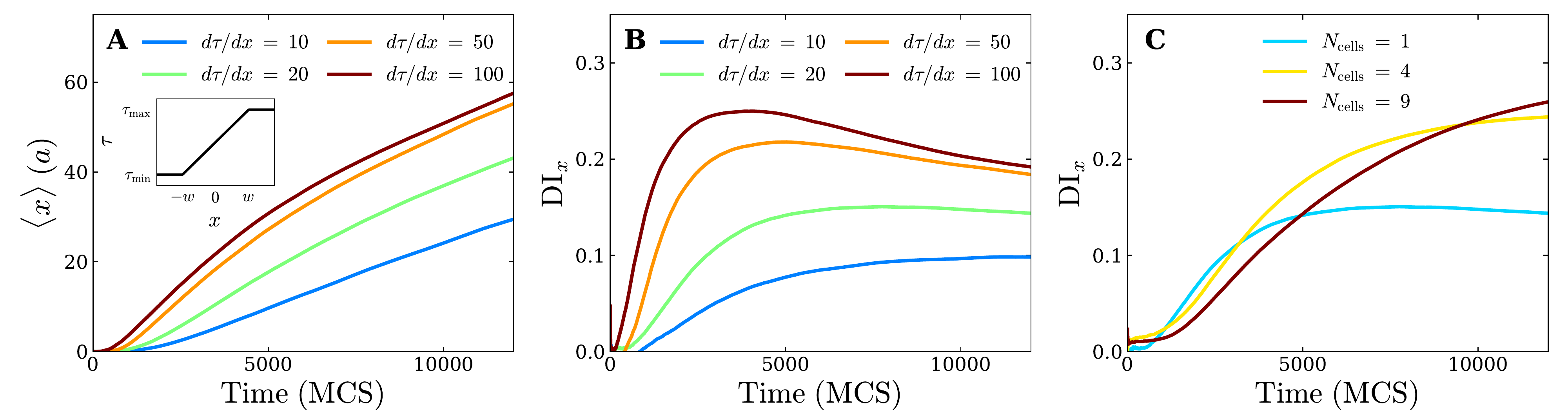}

\caption{Plots of (A) the average displacement in the $x$-direction and (B,C) the $x$-component of the durotactic vector index as a function of time for a single cell (A,B) and fast-aligning different-sized clusters (C) experiencing a linear gradient in persistence time from $\tau_{\mathrm{min}}=200\ \mathrm{MCS}\ (\sim 0.2\mathrm{h})$ to $\tau_{\mathrm{max}}=2000\ \mathrm{MCS}\ (\sim 2.0\mathrm{h})$. Results A and B correspond to different gradients and C to a fixed gradient of $d\tau/dx=20\ \mathrm{MCS}/a$. Gradients are controlled by the width $w$ of the gradient region (see inset). Simulation parameters used: $\gamma=1$ and $\kappa=5$. Averages taken over $10^{4}$ trajectories.}
\label{Single_cell_duro}
\end{figure*}

\noindent In the discussion above, we have demonstrated numerically, and explained theoretically, how mutual velocity alignment can increase the collective persistence of the motion of cell clusters. So far, however, we have assumed the environment of the cell clusters (typically, the ECM) to be homogeneous, and we have included its interactions with the cells only via constant values of the persistence time ($\tau_{\sigma}$), the active speed ($\kappa_{\sigma}$) and the adhesion coefficient ($J_{\sigma,0}$). 
To establish proof-of-principle for the effects of alignment in an inhomogeneous environment, we also extend the CPM simulation setup to include durotaxis (migration in a stiffness gradient); a behavior that may be closely linked to cell persistence. Experimental results suggest that cells on substrates with higher stiffnesses tend to exhibit greater persistence (longer persistence times) \cite{Duro4,Duro5} and simulations of persistently moving point particles have shown that a gradient in persistence time, in itself, is sufficient to generate durotactic motion \cite{Duro2}. 
Moreover, experiments on a larger length scale have indicated that in multicellular settings, entire cell monolayers may exhibit much stronger durotaxis compared to their isolated constituents under the same circumstances; an observation that has in part been attributed to the fact that a larger cellular collective experiences a larger stiffness differential between its leading and trailing edge \cite{Duro1}. This suggests that enhanced durotaxis might also occur in a smaller aggregate like the typical CTC cluster.

Motivated by these experiments, and following the approach in \cite{Duro2}, we model durotaxis by implementing a position-dependent single-cell persistence time $\tau_{\sigma}=\tau (x)$ which, for convenience, is the same for all simulated cells and only depends on their center-of-mass position along the $x$-axis. We then let $\tau(x)$ increase linearly from a minimum value $\tau_{\mathrm{min}}$ to a maximum value $\tau_{\mathrm{max}}$ over a region $x\in [-w,w]$ around the origin, with the cluster (or single cell) center of mass always starting in the middle of the gradient. Beyond the gradient region the parameters remain constant, such that $\tau(x)=\tau_{\mathrm{min}}$ for $x\leq w$ and $\tau(x)=\tau_{\mathrm{max}}$ for $x\geq w$ (see inset \cref{Single_cell_duro}A). This means that the width $w$ effectively controls the steepness of the gradient $d\tau /dx$. The stiffness gradient is always along the positive $x$-direction, and as in most experimental setups the gradient only occupies part of the system \cite{Duro2}, connecting two regions of approximately constant stiffness or persistence time.

To first test whether our implementation of durotactic motion is consistent with earlier simulation work on point particles \cite{Duro2}, we have studied single cell CPM simulations for different gradients $d\tau/dx$ between $\tau_{\mathrm{min}}=200\ \mathrm{MCS}$ ($\sim 0.2\mathrm{h}$) and $\tau_{\mathrm{max}}=2000\ \mathrm{MCS}$ ($\sim 2.0\mathrm{h}$). \Cref{Single_cell_duro}A-B show the calculated $x$-components of the average cell (cluster) displacement $\avg{x(t)}$ and the durotactic vector index $\mathrm{DI}_{x}(t)\equiv \avg{x(t)}/v_{0}t$ respectively. The latter provides the fraction of the average drift velocity of the cell (cluster) along the gradient relative to its absolute speed $v_{0}$ and allows us to quantify the drift up the stiffness gradient \cite{Duro2,Duro6}. Note that $y$-components are not reported, since there is no gradient along this axis and thus no drift. The results are consistent with the fact that a gradient in persistence time suffices to produce a single cell flux towards the stiff region of the domain (positive $\avg{x(t)}$), leading to a form of durotaxis. We also observe an increase (on the investigated time scale) of the drift velocity for increasing gradients (larger values of $\mathrm{DI}_{x}(t)$). Additionally, note that $\mathrm{DI}_{x}(t)$ peaks and afterwards seems to decrease in the long time limit, which is a result of cells leaving the gradient region. A mapping of the retrieved results to the ones obtained for point particles presented in \cite{Duro2} shows that they are also quantitatively the same, making our work fully consistent with literature and extending the point-particle results to cells with a finite area in the CPM.


Having confirmed our implementation of single cell durotaxis, we now proceed by placing different-sized aligned clusters in a fixed gradient $d\tau/dx=20\ \mathrm{MCS}/a$ between the extremes $\tau_{\mathrm{min}}$ and $\tau_{\mathrm{max}}$ (that is, identical environments but different cluster sizes). As one can see in \cref{Single_cell_duro}C, larger clusters indeed show stronger durotaxis, in the sense that the maximal durotactic index is larger for clusters consisting of more cells. We also see that  $\mathrm{DI}_{x}(t)$ takes longer to peak for larger clusters. Comparing cluster behavior to that of single cells, the enhanced durotaxis can be attributed to the enhanced persistence of clusters in combination with the fact that the cluster---simply because it is larger---spans a wider gradient region and thus experiences a larger persistence differential between its front and rear end. As a result, durotaxis of a strongly aligning cluster may also effectively be treated as that of a single, fixed-size particle that moves in an increasingly steep gradient as $N$ increases. 

A prediction that follows from this observation is that increasing the distance between the leading and trailing edge of our model cell clusters, keeping the persistence gradient the same, will generally enhance collective durotaxis. This is indeed what is seen in the experiments reported in \cite{Duro1}, and suggests that even in the absence of long-range force transmission any cluster (with sufficient cell-cell adhesion to maintain cluster integrity) will show enhanced durotactic efficiency as they navigate stiffness gradients.

\section*{Conclusion}
\noindent We have studied and compared the motility of single cells and small cell clusters in the context of tumor cell clusters using a combination of cellular Potts modeling and analytical active-matter theory. CTC clusters have been suggested to pose a far more serious threat than single cells in terms of metastatic potential, and while metastasis is far more involved than motility alone, differences between the ways that small clusters and single cells navigate their environment might play an important part in this striking difference. Our primary aim has been to gain more insight into how the motile behavior of small clusters is affected by their size, and in particular we have focused on whether the effect of cell-cell alignment provides a possible mechanism for enhanced directional motion of clusters.  

We have first carried out CPM simulations to study cell motion in a homogeneous environment. Our single cell simulations show excellent agreement (evidenced by their mean squared displacement) with the persistent random walk, and with experimental results. Moreover, the emergent cell speed and manifested persistence time are directly controlled by model parameters. Extending the simulations to small cell clusters, we have examined the effect of cell-cell alignment by adding a Vicsek-like term to the CPM. Our results demonstrate that alignment enhances the persistence time of cell clusters; the enhancement scales linearly with the number of cells. This allows the cluster to cover more distance than a single cell, which may play some part in its potential to invade the extracellular matrix in the early stages of the metastatic cascade. Within our CPM description, however, this advantage can be suppressed, partly or completely, by adding only one rapidly reorienting cell to the cluster. In addition, we have found that reducing the strength of alignment beyond a critical point results in rapid disintegration of the cluster.

To explain the CPM results, we have proposed a theoretical model in which CTCs are represented by a cluster of fast-aligning active Brownian particles (ABPs). Our analysis reveals that fast velocity alignment increases the persistence time of the ABP cluster, yielding, consistent with the CPM simulations, a linear scaling with the number of particles. The added effect of alignment on the overall cluster mobility is strongest for identical particles, and is also counteracted by adding a rapidly reorienting particle to the cluster.

As a first attempt to investigate the consequences of cell-cell alignment in a more biologically relevant, inhomogeneous environment, we have investigated durotaxis, i.e.\ the migration up a stiffness gradient. Within our CPM simulations we have implemented such a stiffness gradient as a linear gradient in the persistence time. In this scenario, we have shown that in a fixed gradient there indeed exists a durotactic benefit for larger clusters, which may be attributed to the overall persistence differential between the leading and the trailing edge of the cluster and the enhanced cluster persistence due to cell-cell alignment.

Overall, we have shown that, in the presence of generic velocity alignment, single-cell and cluster migration can be significantly different. In particular, enhanced directional migration is exhibited by larger clusters when alignment is fast compared to a typical persistence time in the system. Since these persistence times for living cells are generally on the order of several hours, the condition of rapid alignment may be quite broadly met. 

Our results offer specific predictions for the scaling of both the persistence time, as well as the random motion, of clusters of cells as a function of cluster size. These results fill in a previously uncharted regime between single-cell behavior and large-scale collective motility in confluent cell sheets, a physiologically very relevant regime for which our predictions should be directly observable in experiments.

\section*{Acknowledgments}
\noindent We acknowledge the Netherlands Organisation for Scientific Research (NWO) for financial support through a START-UP grant (V.E.D.).

\appendix

\section{Aligned Cell Cluster Motion in 3D}\label{AppA}
\begin{figure*}[th!]
\includegraphics[scale=0.45]{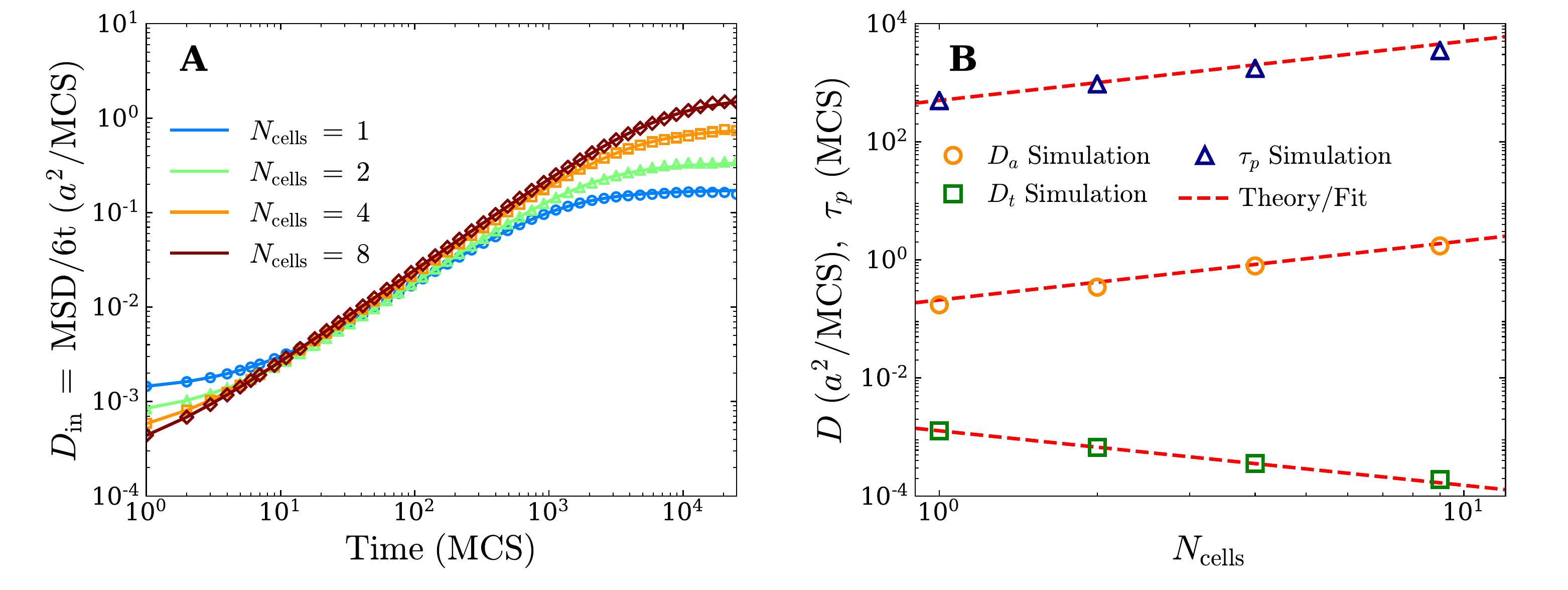}
\caption{(A) Plots of the instant diffusion coefficient $D_{\mathrm{in}} = \mathrm{MSD}/6t$ for fast-aligning ($\gamma=1$) 3D cell clusters consisting of a variable amount of $N_\mathrm{cells}$ identical cells (markers). The results have been fitted with a PRW (lines). (B) Cluster persistence time $\tau_{p}$, active diffusion coefficient $D_{a}$, and thermal diffusion coefficient $D_{t}$ obtained from the fits shown in (A). The obtained values have been compared to or fitted with the derived results of the 2D fast-aligning active Brownian particle theory, i.e.\ \cref{Fast_align_res}. Simulation parameters used: $\kappa=30$ and $\tau = 500\ \mathrm{MCS}$. Data has been obtained by time-ensemble averaging over $20$ trajectories each consisting of $50000\ \mathrm{MCS}$.}
\label{3D_alignment}
\end{figure*}

\noindent To extend the CPM to a 3D system of fast-aligning cells we require new updating rules for the polarity vector $\mv{p}_{\sigma}$, which is now described by the spherical angles $\phi_{\sigma}(t) \in [0,2\pi)$ and $\theta_{\sigma}(t) \in [0,\pi)$: $\mv{p}_{\sigma}=[\sin (\theta_{\sigma})\cos (\phi_{\sigma}),\sin (\theta_{\sigma})\cos (\phi_{\sigma}),\cos(\theta_{\sigma})]$. Discretising the angular Langevin Dynamics for a 3D active Brownian particle we obtain \cite{ABP5}
\begin{equation}
    \begin{split}
        & \theta_{\sigma}(t+\Delta t) = \theta_{\sigma}(t) +  \sqrt{\frac{1}{\tau_{\sigma}}} \Gamma (\Delta t) + \frac{\Delta t}{2\tau \tan (\theta_{\sigma})} ,\\
        \\
        & \phi_{\sigma}(t+\Delta t) = \phi_{\sigma}(t) + \sqrt{\frac{1}{\tau_{\sigma}}} \frac{1}{\sin (\theta_{\sigma})} \Gamma (\Delta t),
    \end{split}
\end{equation}
which in turn, as a result of Vicsek alignment, are extended to
\begin{equation}
    \begin{split}
        \theta_{\sigma}(t+\Delta t) = & \arg_{\theta} \left( \gamma \mv{p}_{\sigma}(t)+\sum_{\sigma^{\prime}} \mv{p}_{\sigma^{\prime}}(t)\right) \\
        & +  \sqrt{\frac{1}{\tau_{\sigma}}} \Gamma (\Delta t) + \frac{\Delta t}{2\tau_{\sigma} \tan (\theta_{\sigma})} ,\\
        \\
        \phi_{\sigma}(t+\Delta t) = & \arg_{\phi} \left( \gamma \mv{p}_{\sigma}(t)+\sum_{\sigma^{\prime}} \mv{p}_{\sigma^{\prime}}(t)\right)\\
        & + \sqrt{\frac{1}{\tau_{\sigma}}} \frac{1}{\sin (\theta_{\sigma})} \Gamma (\Delta t).
    \end{split}
\end{equation}
Here $\arg_{\theta}(\mv{a})$ and $\arg_{\phi}(\mv{a})$ denote the spherical coordinates $\theta$ and $\phi$ of a vector $\mv{a}$ respectively.

Using these updating rules we have calculated the MSDs for 3D fast-aligning ($\gamma=1$) clusters consisting of different numbers of $N_{\mathrm{cells}}$ identical cells. Realising that the MSD of 2D and 3D ABPs are identical up to a change $4D_{t}\rightarrow 6D_{t}$, we have again fitted the results to a PRW. Plots for $D_{\mathrm{in}}=\mathrm{MSD}/6t$ including these fits and the respective fit parameters ($\tau_{p}$, $D_{a}$, $D_{t}$) are shown in \cref{3D_alignment}. Comparing with the 2D results (\cref{Alignment_identical}) we see almost the same behavior, i.e.\ $D_{a}$ increasing linearly with $N_{\mathrm{cells}}$, $D_{t}$ decreasing (almost) linearly with $N_{\mathrm{cells}}$, and $\tau_{p}=N_{\mathrm{cells}}\tau$. This shows that the effect of our Vicsek alignment on the dynamics is qualitatively the same for 2D and 3D CPM cell clusters.

\section{Circular Shape Constraint} \label{AppB}
\noindent As shown in the Hamiltonian $\mathcal{H}$, i.e.\ \cref{Ham}, we model cell-cell attachment via the adhesion coefficient $J_{\sigma,\sigma^{\prime}}$. In particular, by setting the adhesion coefficient between cells $J_{\mathrm{cell-cell}}$ to a sufficiently small value relative to the one between cells and the medium $J_{\mathrm{cell-medium}}$, it becomes energetically more favorable for cells to form a surface with other cells instead of with the medium. However, when this difference becomes too large or the cell-cell adhesion becomes negative, the cells will be able to easily create interfacial area with the other cells, which can lead to a disintegration of the cell shape. This is clearly unphysical behavior. To prevent it from happening, we impose a shape constraint on the cells that forces the cells to have a circular or spherical shape. We can interpret the constraint as a bending rigidity of the cells and formulate it in the form of an energy bias given by \cite{CPM12}
\begin{equation}
\begin{split}
\Delta \mathcal{H}_{r} & = \lambda_{r}  \left( r_{\sigma(\mv{x}_{i})} - \abs{\mv{x}_{i}-\mv{R}_{\sigma(\mv{x}_{i})}} \right) (1-\delta_{\sigma(\mv{x}_{i}),0}) \\
& - \lambda_{r}  \left( r_{\sigma(\mv{x}_{j})} - \abs{\mv{x}_{i}-\mv{R}_{\sigma(\mv{x}_{j})}} \right) (1-\delta_{\sigma(\mv{x}_{j}),0}).
\end{split}
\end{equation}
Here $\lambda_{r}$ denotes the relative strength of the constraint and $r_{\sigma}$ is the preferred radius of cell $\sigma$ so that its area or volume fits precisely in a circle or sphere respectively. The scalar $\abs{\mv{x}_{i}-\mv{R}_{\sigma}}$ denotes the length of the vector that points from the center of mass of cell $\sigma$, i.e.\ $\mv{R}_{\sigma}$, to the location of the candidate site $\mv{x}_{i}$ and can be seen as a local cell radius. Note that the function only applies to cells ($\sigma>0$).

We can explain the form of the energy bias by noting that during each attempt we want to replace the candidate site value $\sigma(\mv{x}_{i})$ by the value of its randomly chosen neighboring site $\sigma(\mv{x}_{j})$. This means that the candidate cell locally retracts at its location $\mv{x}_{i}$, while the neighbor cell locally extends towards $\mv{x}_{i}$. The bias checks whether or not the extension or retraction moves the local cell radius ($\abs{\mv{x}_{i}-\mv{R}_{\sigma}}$) towards or from the preferred radius of the cell $r_{\sigma}$. It then gives a negative energy bias for moves towards the preferred radius, thus making them more favorable. The strength of the energy bias scales with the difference between the local and preferred cell radius; that is, when this difference is large, the cell is more deformed and is therefore more likely to move towards the preferred radius.


\bibliographystyle{apsrev4-1}
\bibliography{all}


\newpage






\end{document}